\documentclass{PoS}
\PoS{PoS(LAT2005)321}
\title{Center flux correlation in SU(2) Yang-Mills theory\footnote{supported in
part by DFG Re 856/4-3}}
\author{K. Langfeld, \speaker{H.~Reinhardt} and G.~Schulze\\
Universit\"at T\"ubingen, Germany\\
E-mail: \email{kurt.langfeld@uni-tuebingen.de},
\email{h.reinhardt@uni-tuebingen.de}}
\abstract{
By using the method of center projection the center vortex part of the gauge
field is isolated and its propagator is evaluated in the center Landau
gauge, which minimizes the open 3-dimensional Dirac volumes of non-trivial
center links bounded by the closed 2-dimensional center vortex surfaces. The
center field propagator is found to dominate the gluon propagator (in Landau
gauge) in the low momentum regime and to give rise to an
OPE correction to the latter of ${\sqrt{\sigma}}/{p^3}$.The screening mass of
the center vortex field vanishes above the critical temperature of the
deconfinement phase transition, which naturally explains the second order 
nature of this transition consistently with the vortex picture. Finally,
the ghost propagator of maximal center gauge is found to be infrared
finite and thus shows no signal of confinement.}
\FullConference{XXIIIrd International Symposium on Lattice Field Theory\\
25-30 July 2005\\
Trinity College, Dublin, Ireland}
\ShortTitle{Center flux correlation in SU(2) Yang-Mills theory}

\usepackage{epsfig}

\newcommand{\bi}{\bigskip}
\newcommand{\no}{\noindent}
\newcommand{\be}{\begin{eqnarray}}
\newcommand{\ee}{\end{eqnarray}}
\newcommand{\hk}{\hspace{0.1cm}}

\newcommand{\hbo}{\hbox to 1 true cm {\hfill } } 

\begin{document}

\no
Although confinement has not yet been
thoroughly understood, several pictures of confinement have been developped, 
which received strong support by lattice calculations in recent years. 
Among these are the dual Meissner effect and the center vortex 
condensation (for a recent review see~\cite{Greensite:2003bk}). 
\bi

\begin{figure}
\begin{minipage}[t]{7cm}
\epsfig{figure=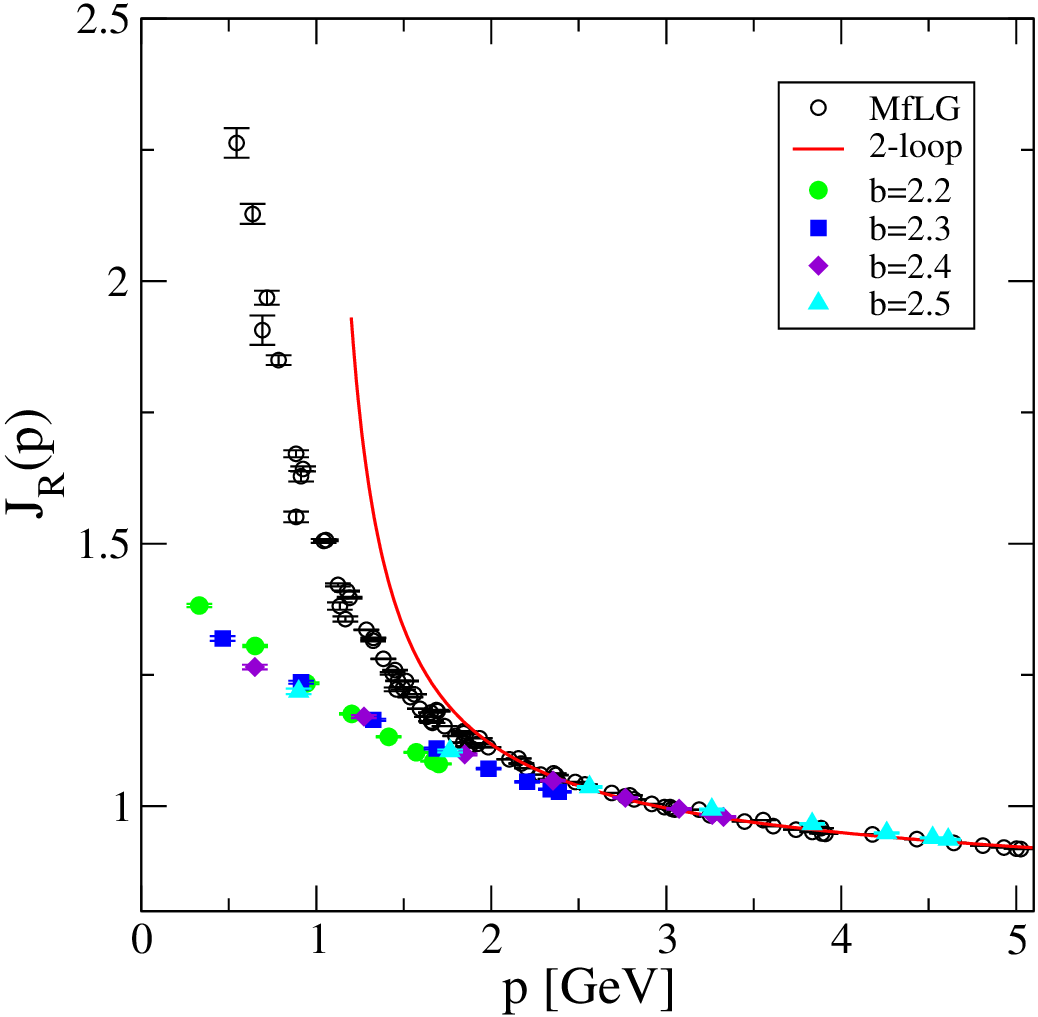,width=.9\textwidth}
\caption{\label{fig:ghost} The ghost form factor (full symbols)
   in MCG as function of momentum transfer. The ghost form 
   factor in Minimal Landau gauge (open symbols, data 
   from~\cite{Bloch:2003sk}). Two loop  perturbation theory (solid line).} 
\end{minipage}
\hspace{0.5cm}
\begin{minipage}[t]{7cm}
\epsfig{figure=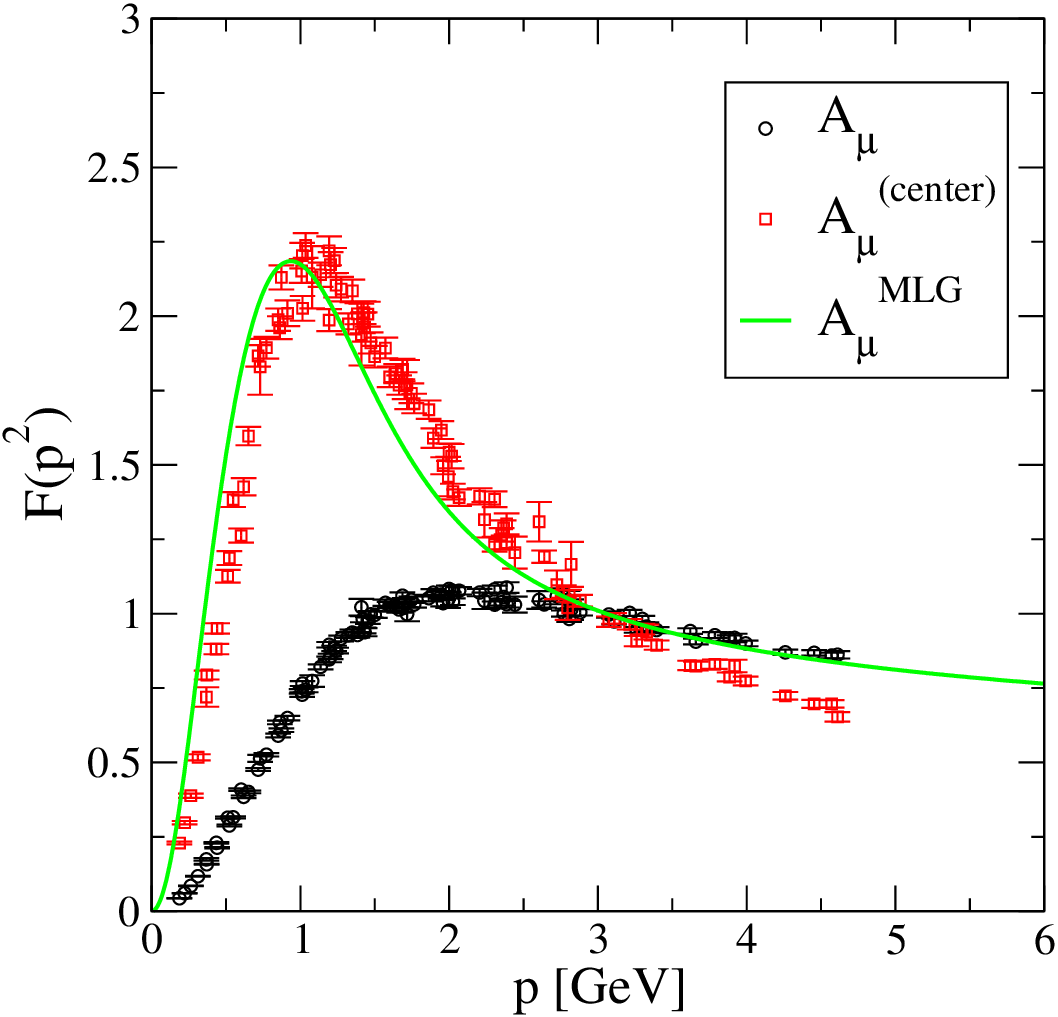,width=.9\textwidth}
\caption{\label{fig:fgluon} The gluon propagator $F(p^2)/p^2$ in MAdL
 gauge as function of the momentum transfer $p$.
  The adjoint gluon form factor 
the gluon 
fields in MAdL gauge (circles), the center field 
form factor 
(squares) and the fit to the  
gluon form factor 
in Minimal Landau gauge (line).}
\end{minipage}\hfill
\hfill

\end{figure}

\no
A different confinement mechanism was proposed by Gribov~\cite{Gribov:1977wm} 
and further elaborated  by 
Zwan\-ziger~\cite{Zwanziger:1991ac}. 
This mechanism is based on the infrared dominance of the field 
configurations near the Gribov horizon, which gives rise to an infrared 
singular ghost propagator, which  is considered to be a signal
of confinement~\cite{Alkofer:2000wg}. 
In Landau gauge this infrared singularity (see fig. \ref{fig:fgluon}) 
disappears, when center vortices 
are eliminated from the Yang-Mills ensemble~\cite{Gattnar:2004bf}. 
Since the infrared singularities are caused by field configurations on the
Gribov horizon, one expects, that center vortices are on the Gribov horizon,
which indeed can be shown to be the case~\cite{Greensite:2004ur}. 
The results of~\cite{Gattnar:2004bf} and~\cite{Greensite:2004ur} show 
that, in Landau and Coulomb gauge, the center vortices are not only on 
the Gribov horizon, but they also dominate the infrared physics. This 
suggests, that the center vortices may be the confiner in any gauge, which is
very plausible since center vortices can, in principle, be defined in a gauge
invariant way and after all confinement is a gauge independent phenomenon. 
\bi

\no
In this talk I will further elaborate on the connection of the two 
confinement scenarios described above, i.e. on the interplay between 
center vortices and ghosts. Using the method of center 
projection, we separate the confining
center degrees of freedom from the remainig (non-confining) coset degrees of
freedom and calculate the associate propagators as well as the corresponding
ghost propagator. We will find, that in 
contrast to the familiar Landau gauge, in the maximum center gauge the ghost
propagator is infrared finite and thus shows no signal of confinement. This
result is not surprising since in this gauge the Faddeev-Popov operator 
does not feel the center part of the gauge field. The confining center 
degrees of freedom constitute the center vortex field. 
We calculate its propagator which is carefully extrapolated to the 
continuum limit. We will find, that the center field propagator
dominates the infrared behaviour of the gluon propagator, while the 
ultraviolet behaviour of the latter is exclusively determined by the coset 
field propagator. However, the center vortex field gives rise to an 
correction to the gluon propagator of the form
$\sqrt{\sigma}/p^3$ which indicates its relevance in the context 
of the operator product expansion. 
\bi

\no
Although the lattice provides a gauge invariant approach to Yang-Mills theory
the transition to the continuum theory is fascilitated by using a gauge, in
which the fields are smooth. The prototype of such a gauge is the wellknown
Landau gauge
\be
\label{1}
\sum_{x, \mu} tr U^\Omega_\mu (x) \stackrel{\Omega }{\longrightarrow } 
\mathrm{max} \hk ,
\ee
We will use various modifications of the Landau gauge to identify the
center vortex content and the remaining coset part of the gauge field as 
will be detailed further below.
\bi

\no
To identify the center vortex content of a gauge field, we use the method of
center projection, 
which is based on the so-called maximal center
gauge (MCG) defined by 
\be
\label{3}
\sum_{x, \mu} \Bigl[ \mathrm{tr} U^\Omega_\mu (x) \Bigr]^2 
\stackrel{\Omega }{\longrightarrow } 
\mathrm{max} \hk ,
\ee
This gauge fixes the gauge group only up to center gauge transformations, i.e.
it
fixes only the coset $SU (2) / Z (2) = SO (3)$ and brings a given link
$U_\mu (x)$ as close as possible to a center element ($\pm1$ for $SU (2))$. This
gauge is just the minimal Landau gauge for the adjoint representation which does
not feel the center.
Once, this gauge is
implemented, center projection implies to replace a link $U_\mu (x)$ by its
closest center element, which is given by 
\be
\label{4}
Z_\mu (x) = \mathrm{sign} \; \mathrm{tr} \,  U^\Omega_\mu (x) \hk .
\ee
The center projected configurations $Z_\mu (x)$ form 3-dimensional volumes of
links $Z_\mu (x) = - 1$, the closed boundaries of which represent the center
vortices. 
We separate the center projected vortices from
the original gauge fields, by writing~\cite{deForcrand:1999ms} 
\be
\label{2A*}
U^\Omega_\mu (x) = Z_\mu (x) \bar{U}_\mu (x) \hk .
\ee

Fig.~\ref{fig:fgluon} shows the ghost form factor of the MCG obtained on $16^4$
and $24^4$ lattices and for $\beta$-values of the Wilson action in the range
$\beta \in [2.15, \, 25]$. At high momenta the ghost form factor of MCG
approaches the one of the minimal Landau gauge but differs drastically from the
latter in the infrared: While the form factor of the minimal Landau gauge is
infrared divergent (what is considered as a signal of confinement), the one of
MCG, which does not feel the center, seems to be infrared finite. This is
consistent with the result obtained in~\cite{Gattnar:2004bf}, that the infrared
divergent behaviour of the ghost form factor in minimal Landau gauge disppears,
when center vortices are removed from the Yang-Mills ensemble, i.e. when the
full links $U^{\Omega}_{\mu} (x)$ (\ref{2A*}) are replaced by their coset parts
$\bar{U}_{\mu} (x)$. 
\bi

\no
The MCG condition (\ref{3}) does not feel the center and accordingly the
Faddeev-Popov operator (and thus the ghost form factor) of MCG depends
only on the coset fields, $\bar{U}_{\mu} (x)$, and consequently does
 not change when
center vortices are removed. Indeed, with respect to the gauge fixing functional
of the MCG (\ref{3}) the center vortices $\{ Z_{\mu} (x) = - 1 \}$ are degenerate
with the vacuum $\{ U_{\mu} (x) = 1 \}$ and for a pure center vortex
configuration $\{\bar{U}_{\mu} (x) \equiv 1 \}$ the Faddeev-Popov 
operator of the MCG
becomes the ordinary Laplacian $(- \Delta)$. Hence the center vortices are not
on the Gribov horizon of the MCG, as opposed to that of the minimal Landau
gauge.
\bi
%

\no
Fig.~\ref{fig:ghost} shows the form factor of the propagator of the gauge field
extracted in the standard fashion from the coset field configurations
$\bar{U}_{\mu} (x)$. 
For large momenta this quantity reproduces the
perturbative result for the full gluon form factor in Landau gauge. 
Like the full 
gluon form factor in Landau gauge, $F_\mathrm{ad} (p^2)$ vanishes 
for $p \to 0$, signaling a mass gap in the excitation spectrum of 
$\bar{A}_\mu (x)$. Finally, $F_\mathrm{ad} (p^2)$ deviates essentially 
from the  gluon form factor of minimal Landau gauge in the
intermediate momentum regime.
\bi

\begin{figure}
\begin{minipage}[t]{7.5cm}
\center\epsfig{figure=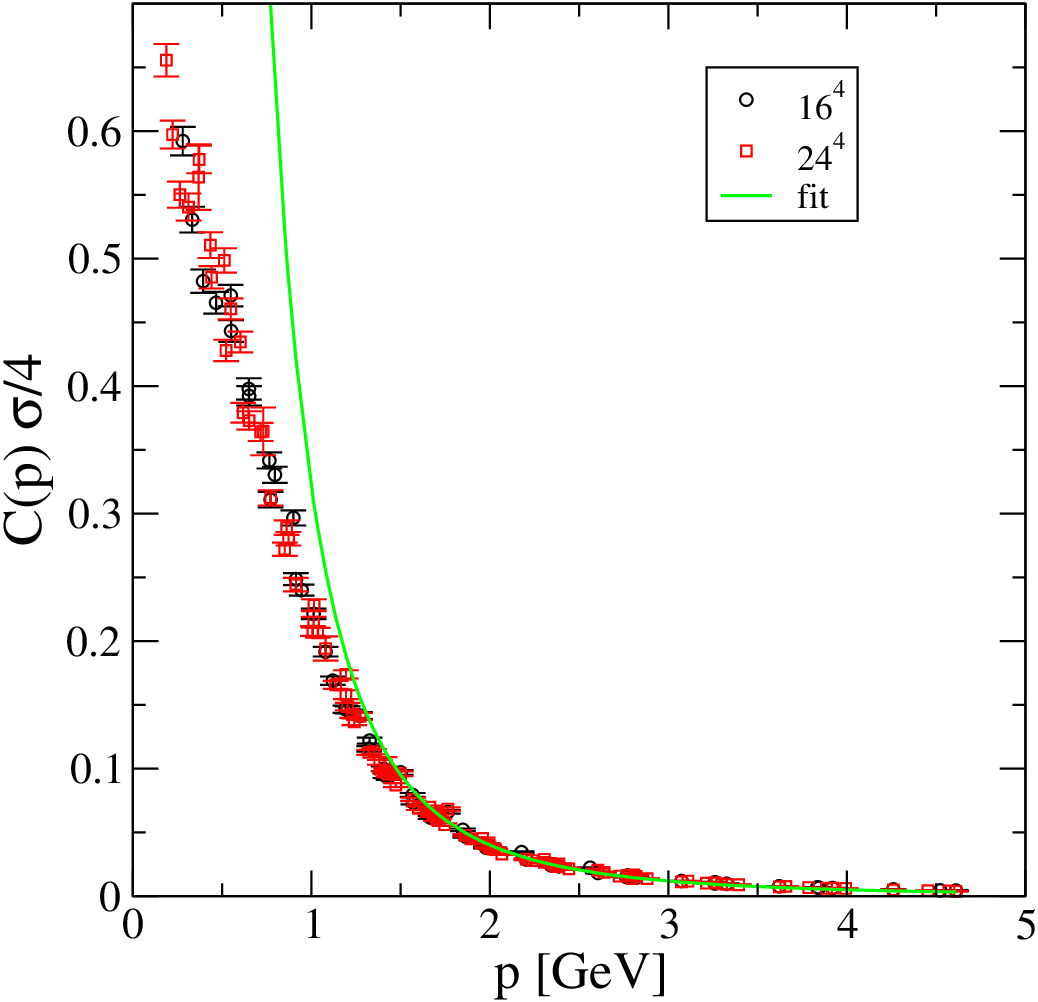,width=.9\textwidth}
\caption{\label{fig:cenp} Center field correlation function as function 
   of momentum for two different lattice sizes.} 
\end{minipage}\hfill
 \begin{minipage}[t]{7.5cm}
 \center\epsfig{figure=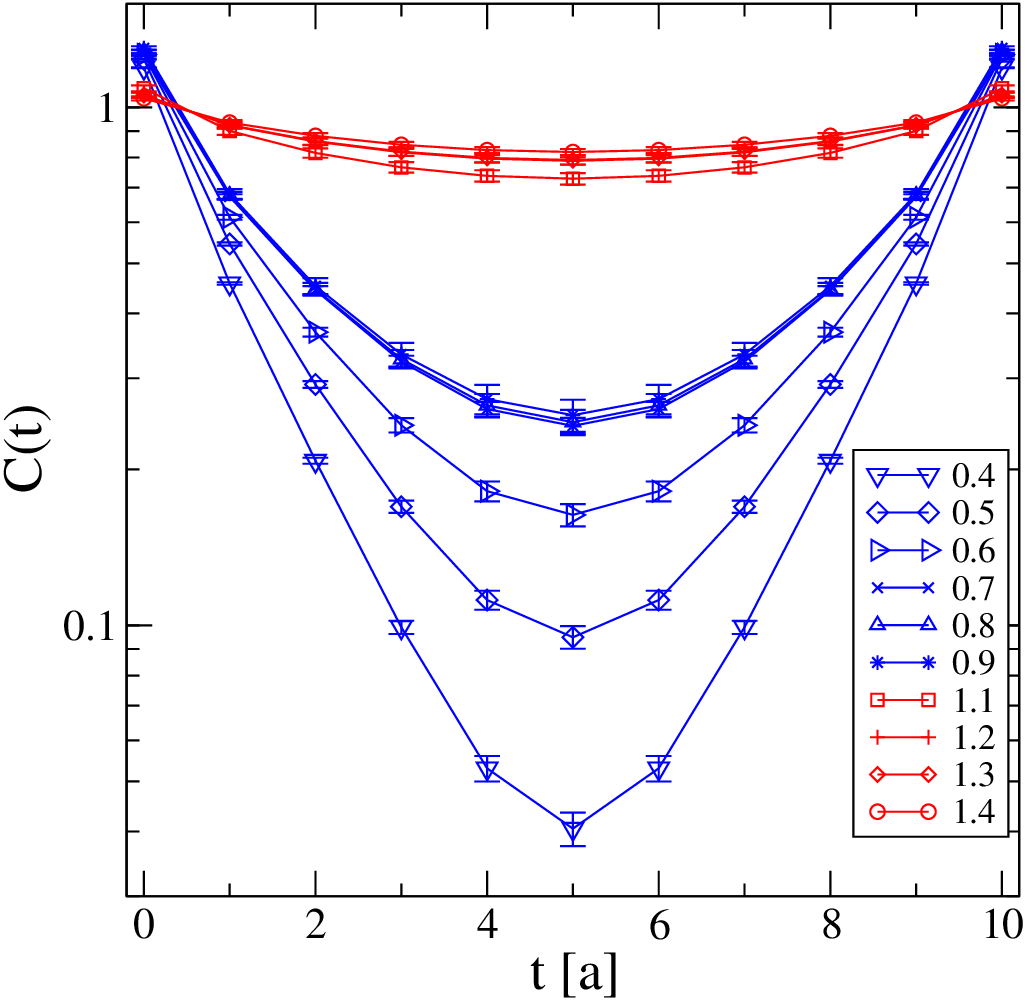,width=.9\textwidth}
\caption{\label{fig:ct} Center field correlation function as function  
  of Euclidean time for several temperatures (in units of the critical 
  temperature).} 
\end{minipage}
\end{figure}
Consider now the center
projected configurations $\{ Z_{\mu} (x) \}$ (\ref{4}). 
Each link $Z_\mu
(x)$ defines an elementary cube on the dual lattice and the total number of
non-trivial center links $Z_\mu (x) = - 1$ defines  open
hypersurfaces $\Sigma$ bounded by closed center vortex surfaces $\partial
\Sigma$.

\no 
Introducing the characteristic function of the hypersurfaces $\Sigma$ by 
\be
a \chi_{\mu} (\Sigma, x) = \frac{1}{2} \, (1 - Z_{\mu} (x)) = \left\{
\begin{array}{ccl}
1 & , & x \in \Sigma \\ 0 & , & x \notin \Sigma
\end{array} \right. \hk ,
\ee
one can define a vector field 
\be
{\cal A}_\mu (\Sigma, x) = \pi \tau_3 \chi_\mu (\Sigma, x) 
\ee
satisfying $ \exp \{ ia {\cal A}_\mu (x) \} = Z_\mu (x) $.

We also use the analogue of the Landau gauge for the center
projected fields 
\be
\label{12}
\sum_{x, \mu} Z_\mu (x) \stackrel{\Omega _2}{\longrightarrow } 
\mathrm{max} \,  \hk , 
\ee 
where $\Omega _2 $ is a $Z (2)$ gauge transformations. 
This
gauge condition, eq. (\ref{12}), minimizes the number of $Z_\mu (x)
= - 1$ links and thus minimizes the volume of 
the open 3-dimensional hypersurfaces $\Sigma$. Since, the
minimal open hypersurfaces are completely determined by their
boundary\footnote{: For topologically non-trivial
 surfaces $\partial \Sigma$ there may 
be more than one (relative) minimal open surfaces $\Sigma$. Nevertheless 
all these minimal surfaces are completely determined by their common 
boundary $\partial \Sigma$.}, 
it is expected that they scale properly towards the continuum limit,
if the vortex surfaces $\partial \Sigma$ do. 
The probability ${\cal P}$ that a given link element $Z_\mu (x)$ 
is negative should scale
in the continuum limit $a \rightarrow 0$ 
as ${\cal P} \; = \; \kappa \; a$, 
with $\kappa $ being independent of $a$. This is indeed confirmed by 
lattice calculations~\cite{Kovalenko:2004xm}.  
Our numerical calculations yield $\kappa \approx 0.26 (1) \sqrt{\sigma}$.
\bi

%
%
\no
Consider now the connected center field correlation function 
\be
\label{18}
a^2 c_{\mu \nu} (x-y) = \langle Z_\mu (x) Z_\nu (y) \rangle 
\; - \; \langle Z_\mu (x)
\rangle \langle Z_\nu (y) \rangle \hk . \\
= a^2 \, \frac{4}{\pi^2} \, \bigg[ \langle {\cal A}_{\mu} (x) \, {\cal
A}_{\nu} (x) \rangle - \langle {\cal A}_{\mu} (x) \rangle \, \langle {\cal
A}_{\nu} (x) \rangle \bigg] \nonumber
\ee 
The factor $a^2$ was introdcued for later convenience. Although 
$Z_\mu (x)$ is an integer valued field, the propagator $c_{\mu \nu} (x-y)$, 
emerging from an ensemble average, appears to be smooth 
and transverse: $ \partial _\mu \; 
c_{\mu \nu} (x-y) \; = \; 0 $. 

\no
One of our important findings is that the 
center field correlation function $C_{\mu \nu }(p)$ 
is independent of the lattice spacing. The consequences are two-fold: 
Firstly, the propagator  $C_{\mu \nu }(p)$ is {\it not} subjected to wave 
function renormalization. Secondly, $C_{\mu \nu }(p)$ behaves as a genuine 
gluonic correlation function with mass dimension two. 

\vskip 0.3cm
Our numerical results for the propagator 
$ C_{\mu \nu }$ in momentum space are shown in figure~\ref{fig:cenp}. 
Most striking is that the high momentum tail is well fitted by the power 
law (see figure~\ref{fig:cenp}) 
\be 
C(p) /4 \; = \; \frac{ 3.7(1) \; \sqrt{\sigma } }{ p^3 } \; , 
\hbo p \ge 2 \, \mathrm{GeV} \; . 
\label{eq:fit} 
\ee 
This implies that the center field correlator is sub-leading in the 
high momentum regime compared with the perturbative correlator (see also 
figure~\ref{fig:fgluon}). The center field correlator, 
however, contributes to the operator product corrections. Hence, 
the center ``background'' field could serve a natural explanation for 
the condensates entering the operator product expansion. 
Finally note that $ F^\mathrm{cen}(p^2) $ is enhanced in the low momentum 
regime, where it has the same shape as the gluon  
form factor in minimal Landau gauge.

\vskip 0.3cm
Let us finally consider the deconfinement phase transition at 
high temperatures. This transition is well understood in the vortex picture 
where it appears as vortex depercolation
transition. 
Since the transition is of 2nd order, it is accompanied 
by the occurrence of a massless excitation. However, it is known that 
neither the gluonic mass gap nor the color singlet states hardly change
significantly and thus cannot be identified with the 
excitation that becomes massless at the transition. 
Because of the success of the vortex picture in 
describing this transition, one might suspect that the center field 
correlator contains the desired information. We have therefore studied 
the correlator $\sum _{\vec{x}} c(t, \vec{x})$ as function of $t$ 
for several temperatures. 
Simulations have been carried out using a $30^3 \times 10$ lattice. 
$\beta $ was varied from $1.91$ to $2.69$ to adjust the temperature.
The result is shown in fig. 4. At temperatures $T$ below the 
critical temperature ($T_c$) we find an exponential decrease of the 
correlator $C(t)$, while the correlation is compatible with a power-law 
for $T>T_c$.


\end{document}